\documentclass[twocolumn,letterpaper]{IEEEAerospaceCLS}  

\usepackage[]{graphicx}    
\usepackage{chemformula}   
\usepackage{isotope}       
\newcommand{\ignore}[1]{}  

\begin{document}
\title{Enabling Ice Core Science on Mars and Ocean Worlds}

\author{%
Alexander G. Chipps\\
School of Mechanical Engineering\\
Georgia Institute of Technology\\
Atlanta, GA 30332\\
achipps3@gatech.edu	
\and 
Cassius B. Tunis\\
School of Aerospace Engineering\\
Georgia Institute of Technology\\
Atlanta, GA 30332\\
ctunis3@gatech.edu
\and
Nathan Chellman\\ 
Desert Research Institute\\
2215 Raggio Parkway\\
Reno, NV 89512\\
nathan.chellman@dri.edu
\and
Joseph R. McConnell\\
Desert Research Institute\\
2215 Raggio Parkway\\
Reno, NV 89512\\
joe.mcconnell@dri.edu
\and
Bruce Hammer\\
University of Minnesota\\
Center for Magnetic Resonance Research\\
2021 6th Street SE\\
Minneapolis, MN 55455\\
hammer@umn.edu
\and
Christopher E. Carr\\
School of Aerospace Engineering \&\\
School of Earth and Atm. Sciences\\
620 Cherry St NW, Room G10\\
Atlanta, GA 30312\\
cecarr@gatech.edu
\thanks{\footnotesize 978-1-6654-9032-0/23/$\$31.00$ \copyright2023 IEEE}              
}

\maketitle

\thispagestyle{plain}
\pagestyle{plain}

\maketitle

\thispagestyle{plain}
\pagestyle{plain}

\begin{abstract}
Ice deposits on Earth provide an extended record of volcanism, planetary climate, and life. On Mars, such a record may extend as far back as tens to hundreds of millions of years (My), compared to only a few My on Earth. Here, we propose and demonstrate a compact instrument, the Melter-Sublimator for Ice Science (MSIS), and describe its potential use cases. Similar to current use in the analysis of ice cores, linking MSIS to downstream elemental, chemical, and biological analyses could address whether Mars is, or was in the recent past, volcanically active, enable the creation of a detailed climate history of the late Amazonian, and seek evidence of subsurface life preserved in ice sheets. The sublimation feature can not only serve as a preconcentrator for in-situ analyses, but also enable the collection of rare material such as cosmogenic nuclides, which could be returned to Earth and used to confirm and expand the record of nearby supernovas and long-term trends in space weather. Missions to Ocean Worlds such as Europa or Enceladus will involve ice processing, and there MSIS would deliver liquid samples for downstream wet chemistry analyses. Our combined melter-sublimator system can thus help to address diverse questions in heliophysics, habitability, and astrobiology.
\end{abstract}

\tableofcontents

\section{Introduction}
Ice deposits on Earth provide an extended record of volcanism, cosmic events, and subsurface life. On Mars, records entrained in polar and mid-latitude ice may extend as far back as tens to hundreds of My \cite{Bramson2017,Levy2021}, compared to only a few My (and less than one million years contiguous) on Earth \cite{Bergelin2022}. Martian ice investigations would provide a significant extension of our knowledge from Earth ice analysis, and could serve as a precursor to the exploration of icy bodies such as Enceladus and Europa. Such ice, found in deposits more than 100 meters thick in mid-latitudes \cite{Dundas2018}, may also represent an important resource for human exploration.

The layers present within ice cores record historical precipitation (e.g., snow), aerosols, and trap atmospheric gases as the overburden ice accumulates and the pore space is closed off (around 50-80 m deep on Earth).

One common strategy for reducing ice core contamination is to physically remove the outer part of a core, as was done for a recent study assessing the contamination of Arctic ice cores as a measure of the potential of forward contamination during future life detection missions \cite{Coelho2022}. This process is labor intensive and challenging to automate. Melters, which heat an ice core or section of an ice core (ice stick) above its melting temperature at the base, offer an alternative that facilitates automatic processing, and can reduce contamination by permitting automatic disposal of the meltwater from the outer part of the core as described in the next section.

Melters permit sampling of these layers at high resolution, including by coupling meltwater to real-time downstream analyses such as inductively-coupled plasma mass spectrometry (ICP-MS) for elemental abundance \cite{McConnell2002}, laser spectrometers for gas (e.g., methane) detection \cite{Rhodes2013}, black carbon or other compounds \cite{McConnell2007}, or by collection of fractions for additional analyses \cite{Santibanez2018}.

Thus, we propose and demonstrate initial work toward a compact Melter-Sublimator for Ice Science (MSIS) instrument that would enable convenient ice core processing on Mars. MSIS is built upon an extensively validated melter design \cite{McConnell2002,McConnell2007,Rhodes2013,Santibanez2018,Arienzo2019}, and it has a compact and thermally-efficient design to meet the requirements of planetary exploration (size, mass, power, etc.). Through a combination of in-situ and sample return processes, pairing an MSIS-processed sample with downstream chemical analysis instruments may assist in several potential use cases:

\begin{itemize}
  \item [1)] Addressing when and where Mars was last volcanically active. \\
  \item [2)] Confirming and expanding the record of space weather events over a timescale not possible with Earth ice.\\
  \item [3)] Characterizing the inventory of organic molecules on Mars in support of the search for life beyond Earth.
\end{itemize}

In this paper, we describe the MSIS instrument concept, design, and current performance. We also discuss the limitations and future development goals for MSIS, as well as the applications of MSIS to the potential use cases proposed above.

\section{Melter-Sublimator for Ice Science}

\subsection{Concept of Operations}
\begin{figure*} 
    \centering
    \includegraphics[width=7in]{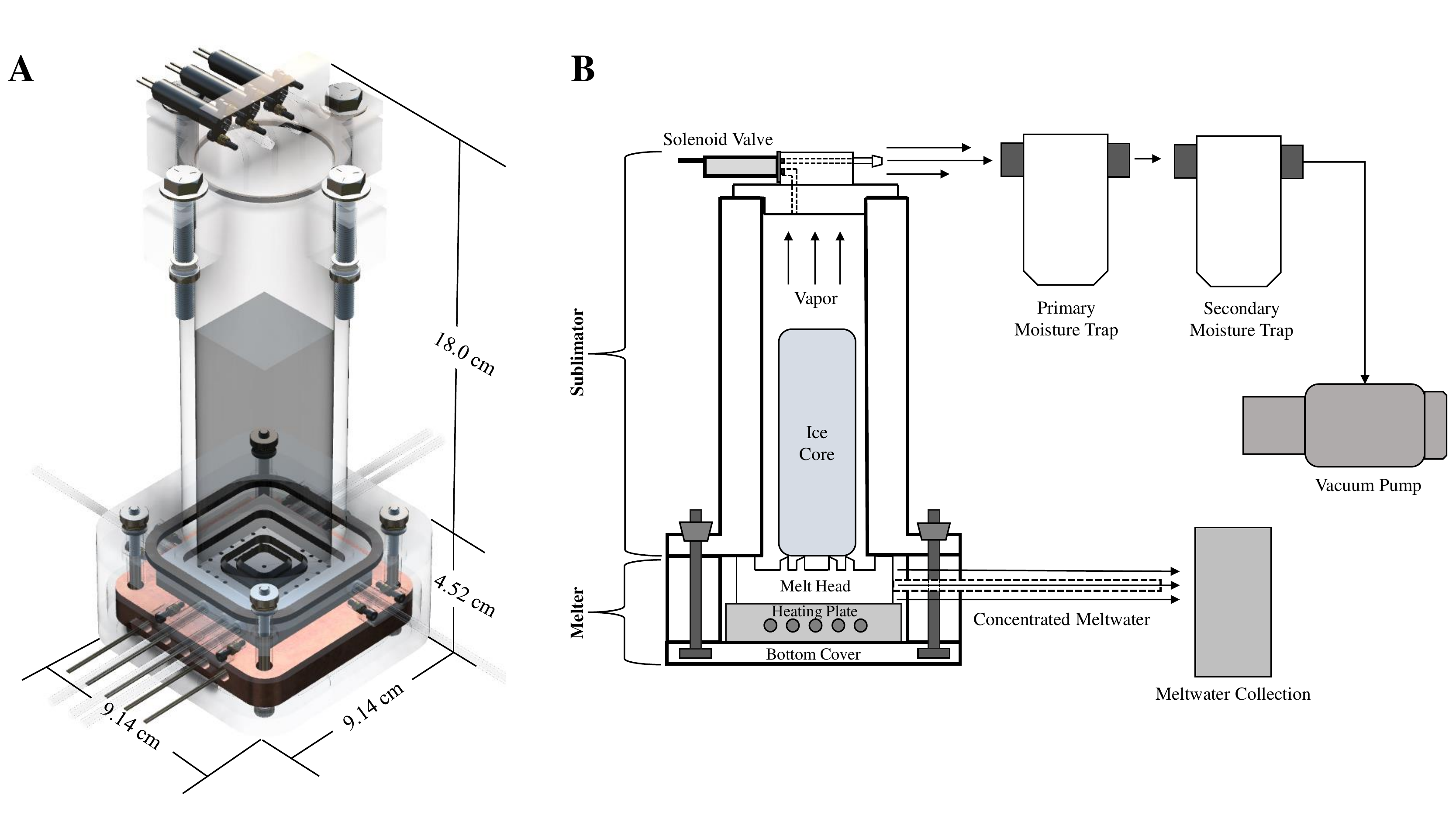}
    \caption{\bf{\label{fig_Overview}Melter-Sublimator for Ice Science (MSIS) Overview. (A) CAD model. (B) Cross-sectional view illustrating both melting and sublimation modes of operation.}}
\end{figure*}

The MSIS instrument can be best described as having two main components: Melter and Sublimator (Figure \ref{fig_Overview}).

On Earth, drilling, extraction, and handling processes result in contamination of an ice core’s outer surface \cite{tr:mars-ice-2021}. Accessing pure, uncontaminated regions of the core and rejecting contaminated areas is a requirement for some downstream (elemental, chemical, biological) analyses \cite{Osterberg2006}. Thus, like the melter design from McConnell et al. \cite{McConnell2002,McConnell2007,Rhodes2013,Santibanez2018,Arienzo2019}, the MSIS Melter component divides an ice stick into three distinct cross-sectional areas as it melts, providing individual access to the inner, middle, and outer core sections.

MSIS is currently designed to accommodate 3.3 cm x 3.3 cm ice sticks, following \cite{McConnell2002}, but in principle could be adapted to any core cross section. Using the three-ring design adopted from \cite{McConnell2002}, water from the outer 70\% of the core is discarded under ideal operating conditions. The inner, most pristine 10\% is collected separately from the middle 20\% to facilitate downstream analysis applications requiring different levels of purity and volume flow rate. The geometry of these rings can be redesigned to suit any specific set of downstream flow requirements.

The Sublimator component, sitting atop the Melter, enables preconcentration of material for in-situ analyses or Earth return of residues, avoiding the complexities involved in storing and returning large samples to Earth. MSIS will be able to preconcentrate organic samples through near-full sublimation then controllably melt a reduced sample, allowing for sensitive in-situ characterization of non-volatile materials such as organics.

\subsection{Melter Design \& Implementation}
\begin{figure*} 
    \centering
    \includegraphics[width=7in]{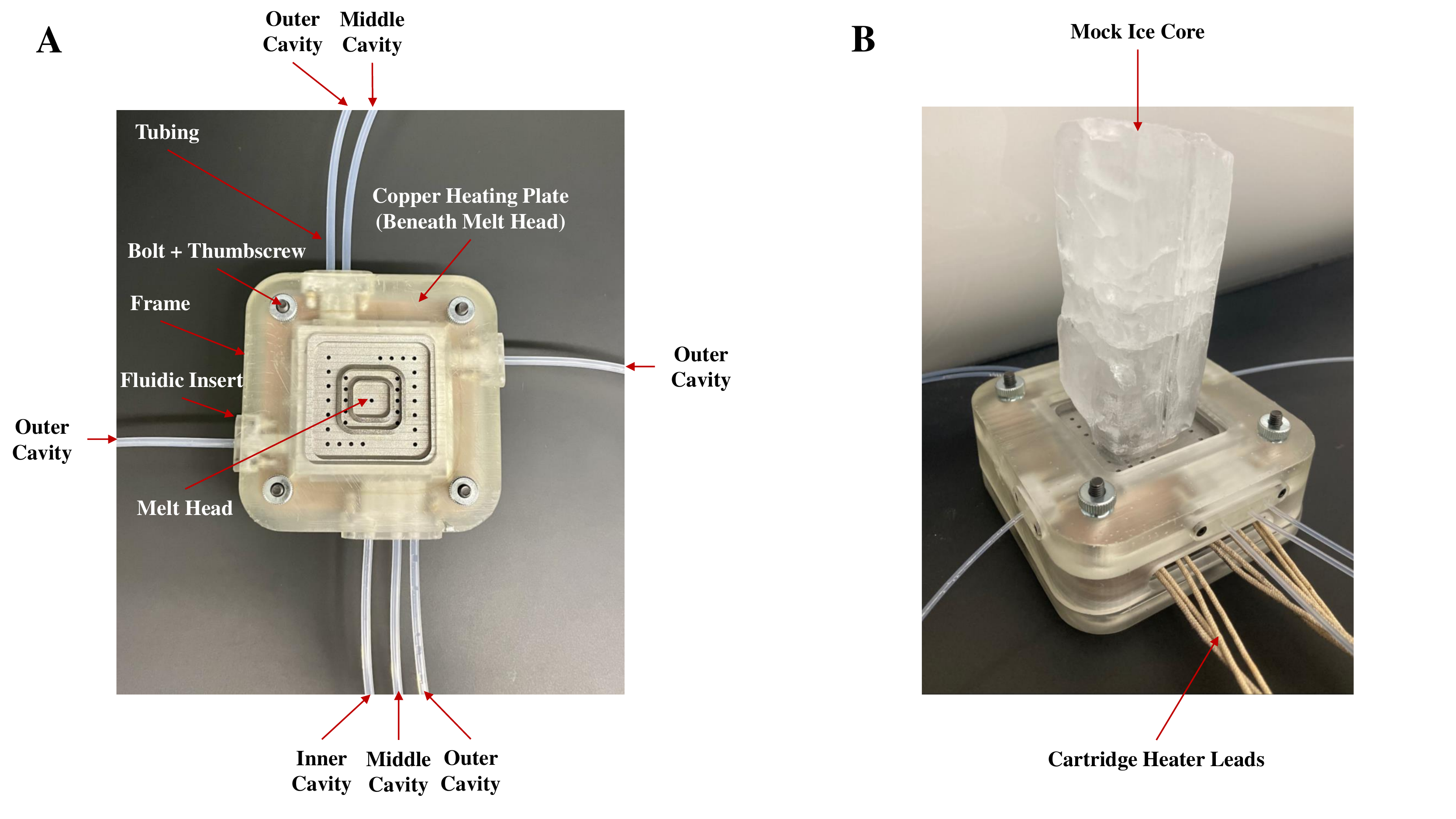}
    \caption{\bf{\label{fig_MelterDesign}(A) Top view of Melter component. Cartridge heaters not shown. (B)  Melter component melting lab-made ice core on benchtop.}}
\end{figure*}

The Melt Head is a stainless steel, 3D-printed structure that sits compressed against a heated copper plate with a layer of thermal paste at the interface. The Melt Head has two ~3.5-mm-tall partitioning walls that separate the ice stick as it melts into the inner, middle, and outer Melt Head cavities (Figure \ref{fig_MelterDesign}). Copper was not used for these partitioning walls because it would be chemically reactive. Also, a stainless steel Melt Head does not permit trace metal-clean operation like a silicon carbide one would; however, stainless steel is an inexpensive compromise for our initial testing. Melt Head cavity dimensions were adapted from the dimensions of the McConnell et al. design \cite{McConnell2002,McConnell2007,Rhodes2013,Santibanez2018,Arienzo2019}, with the outer cavity having a large enough area to melt through a 3.3 cm x 3.3 cm ice stick. Within each cavity are drainage holes that lead into the fluidic channels and to the outlet ports on the sides of the Melt Head.

The Frame and Bottom Cover are Formlabs Clear resin-printed structures that compress the Melt Head to the Heating Plate via bolts and thumb nuts. The Frame also serves as a mounting point for the Fluidic Inserts. The Melt Head is press-fitted by its four corners into the Frame and held only at those locations. Gaps were designed in the Frame to separate it from the Melt Head and the Heating Plate, minimizing heat transfer from these components to the plastic; This is for safe handling of the MSIS system and to conduct as much heat directly from the Heating Plate to the Melt Head as possible. For similar reasons, the Bottom Cover minimizes its contact with the Heating Plate while maintaining structural integrity.

The Fluidic Inserts interface with the outlet ports of the Melt Head. They were designed to mimic nuts commonly used with fluidic tubing and ferrules. There are four Fluidic Inserts: two host 1 fluid line, one hosts 2 fluid lines, and another hosts 3 fluid lines. Fluidic tubing in this design is $1/8$-in-outer diameter (OD), $1/16$-in-inner diameter (ID), fluorinated ethylene propylene (FEP) tubing from IDEX Health \& Science. A custom design for hosting the fluidic tubing was preferred over a commercial product because it can easily be adapted for various design configurations and tubing sizes. Threaded inserts and bolts are used to mount the Fluidic Inserts to the Melter Frame. When tightened down, the Fluidic Inserts compress the tubing with custom ferrules against the Melt Head so that meltwater can be transported from the Melter component to downstream devices via the tubing. O-rings between each custom ferrule and the Melt Head add another measure in preventing meltwater leakage in the system.

Four pumps were added along the fluid lines to provide additional control over the flow rate and to prevent the meltwater from overflowing beyond the Melt Head. Maintaining a higher flow rate at the outer cavity and lower flow rates in the inner two cavities is important for reducing contamination of the inner cavities \cite{Osterberg2006}. Two small, 3 V diaphragm pumps yield higher flow rates and are appropriate for draining the outer cavities, which nominally send contaminated meltwater to waste. Peristaltic pumps are preferred for the middle and inner cavities because the meltwater does not contact the pump internals, preventing sample contamination. One peristaltic pump is used for the middle cavity drain lines, and another for the inner cavity. Each Melter pump can be powered by varying voltages for different flow rates.

The Heating Plate is a $1/2$-in-thick piece of copper manufactured to hold up to five, $1/4$-in-diameter cartridge heating elements. The interfaces between the cartridge heaters and the Heating Plate are covered with thermal paste to lower any contact resistance. Copper was selected as the Heating Plate material because it has high thermal conductivity, $k \approx 385\text{ }\frac{W}{m\cdot K}$, while having a relatively reasonable price compared to more thermally conductive materials like silver or diamond. The MSIS Heating Plate is much thinner than that of other melter designs, meaning the thermal load is lower for the cartridge heaters; This allows for increased thermal efficiency in our instrument. Also, the Heating Plate tends to oxidize with use, so semi-frequent sanding of the surface is done to reduce the thermal load.

\begin{figure*} 
    \centering
    \includegraphics[width=7in]{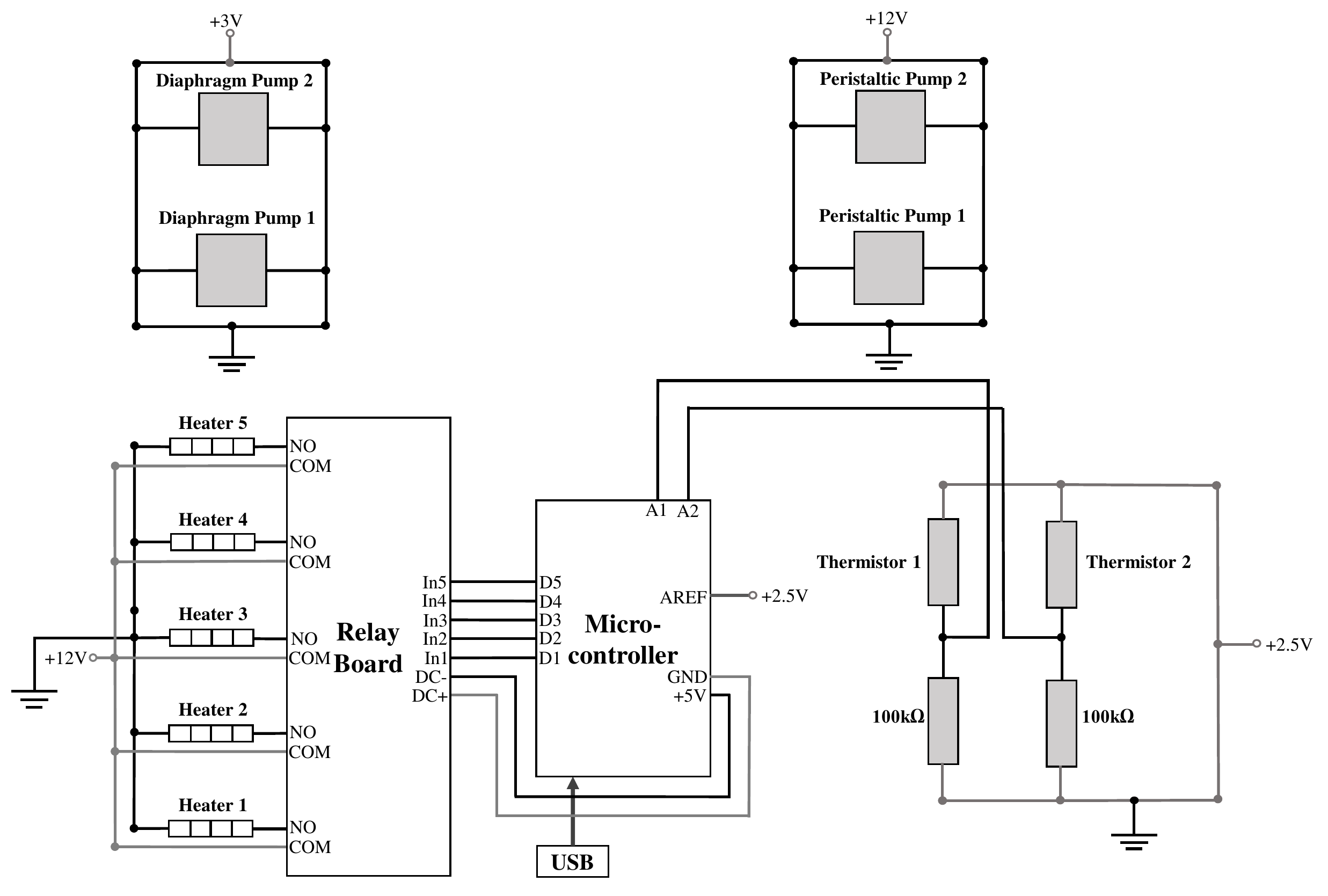}
    \caption{\bf{\label{fig_MelterElectronics}Melter Electronic Wiring Diagram}}
\end{figure*}

We now describe the electrical system that controls the Melter (Figure \ref{fig_MelterElectronics}). Inside the Heating Plate sits the $1/4$ in DC cartridge heaters, powered with an external power supply up to 24 V, which is expected on a typical spacecraft bus. In 12 V operation, each of the five cartridge heaters produces 14 W, therefore the theoretical maximum power supplied to the melter system is 70 W; 12 V was selected for operating the cartridge heaters in the lab, as 70 W of power was adequate for MSIS benchtop testing, and working with lower voltages is good safety practice. Per the slim Heating Plate, the cartridge heaters lie horizontally, in contrast to their vertical orientation in other melter designs \cite{Osterberg2006}. A NOYITO 5 V relay board is used to switch the cartridge heaters via an Arduino Uno microcontroller. Each relay can switch loads up to 10 A and 30 VDC. The relay board in this design was selected because it offers low and high-level triggering, accepts serial commands, and has enough channels to switch five cartridge heaters. 

Temperature is monitored and controlled by a closed-loop Proportional-Derivative (PD) system that relies on temperature measurements taken by two thermistors mounted to the Melt Head (Figure 5). Each thermistor is first in series with a 100 k$\Omega$ resistor and powered by 2.5 V. As the temperature at the thermistor rises, its internal resistance is lowered, and the Arduino microcontroller can measure a larger analog voltage. The 10-bit analog-to-digital converter in the microcontroller processes this analog signal using a lookup table, comparing a given voltage value to an experimentally determined range of voltages versus temperatures. 

A temperature setpoint is specified in the microcontroller. The control system evaluates its current temperature versus the setpoint and begins heating. Then, the PD controller detects when the current rise in temperature is sufficient to turn off the heaters, at which point the control system maintains the temperature setpoint with varying bursts of heat. This PD controller is adaptable to different setpoints and starting temperatures.

\subsection{Sublimator Design \& Implementation}

\begin{figure} 
    \centering
    \includegraphics[width=3.25in]{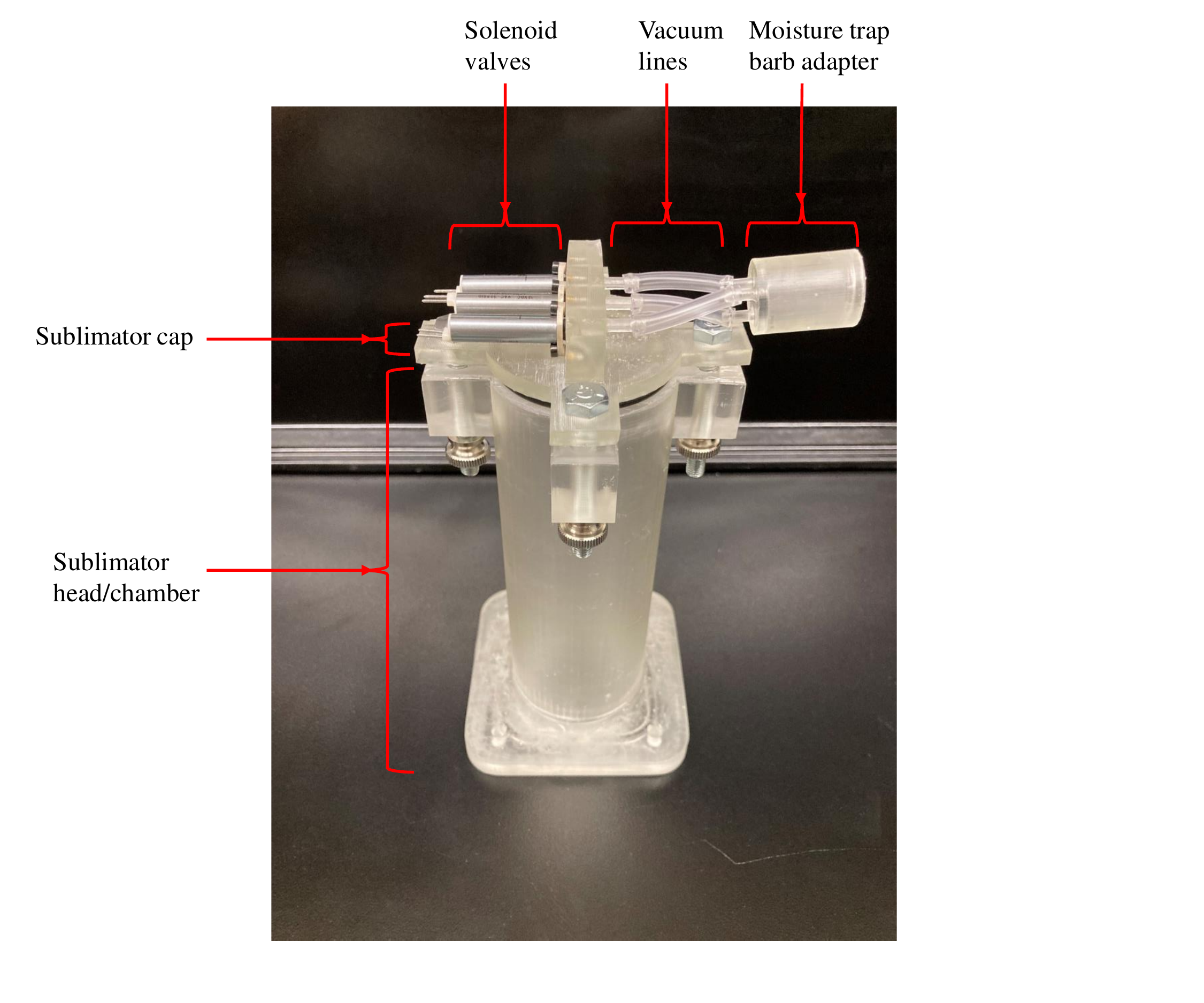}
    \caption{\bf{\label{fig_Sublimator}Sublimator hardware, excluding relay board and electronics.  Sublimator pictured without Melter.}}
\end{figure}

The Sublimator component of MSIS (Figure \ref{fig_Sublimator}) consists of a resin-printed cylindrical sublimation chamber that adapts to the top of the Melter component. The Sublimator is sealed against the Melter by a face-seal O-ring. This allows containment of the ice sample, whether it be fragments or a full ice stick. The current design can house an ice stick measuring 3.3 cm $\times$ 3.3 cm at the base and a height of up to 15 cm. The sublimation chamber is sealed at the top by a cap that screws into the chamber, and three horizontally mounted Lee Company (Westbrook, Connecticut) solenoid valves are attached. The solenoid valves expose the ice core to vacuum when opened. Valve opening is controlled by three solid-state relays connected to the Arduino microcontroller. Through serial command, the operator specifies the number of valves they wish to open, depending on the desired outflow. A CoolCube 50R “hit-and-hold” circuit is placed in series between each relay and each valve. The valves require 12 V to actuate but do not require this voltage to remain actuated. The “hit-and-hold” circuitry reduces the operating voltage to 6 V to prevent overheating while keeping the valve open for extended periods (e.g., during sublimation).

The outlet to the solenoid valves extends into barb connectors to allow the channeling of water vapor into a common moisture trap, which consists of a sealed basin with a cylindrical water and particulates filter. This helps protect the downstream vacuum pump to which the vapor is being vented. Such an adapter may not be necessary in a low-pressure atmosphere, as on Mars, where additional vacuum may be superfluous depending upon the local pressure. 

In its operation, the Sublimator would reduce the pressure in the sublimation chamber to below the triple point of water and raise the temperature of the Melt Head to 0$^{\circ}$C (Figure 5). To simulate Martian conditions more accurately, the Sublimator would reduce the pressure to 1 Torr. Lower pressures can be attained with strong vacuum sources, but pressures significantly lower than 0.75 Torr can negatively impact the sublimation rate by hindering convective heat transfer into the ice \cite{hartmuller2019demonstration}. Higher sample temperatures also favor faster sublimation rates; however, we must maintain a lower temperature at the Melt Head throughout the initial exposure to vacuum to prevent premature melting \cite{hartmuller2019demonstration}. Since sublimation is endothermic, the temperature of the ice stick will decrease throughout the process. Our control over the Melt Head temperature allows us to set temperatures appropriate for any state of the ice stick during the sublimation process.

Depending on the application, the Sublimator does not require all the functionalities of the Melter. The Sublimator component is designed to both interface with the entire Melter component or only the Heating Plate. This allows for the use of a thermal plate inside a vacuum chamber to heat a sample through the Melter Heating Plate, thereby eliminating the need for a cartridge heater-relay setup where space is limited (Figure 5). For other applications, only a reduced solid sample may be desired, so melting may not be necessary.

\subsection{Proof of Principle Testing}
\subsubsection{Thermal Efficiency of Melting}

Thermal efficiency of the MSIS Melter component is desired to optimize power usage onboard a spacecraft or in an extraterrestrial environment. Cold environments, such as the Antarctic ice sheet, Mars, or Europa will work against the ease of melting, both demanding more power from electronics and potentially lowering the melt rate. To improve thermal efficiency, we maximize the amount of conductive surface contact between the cartridge heaters and the Melt Head and minimize contact with all other parts. The heaters can individually attain 100$^{\circ}$C in 60 seconds when powered by 12 V. This heating time is sufficiently fast to justify using 12 V as opposed to 24 V to power the cartridge heaters, which would otherwise increase the power consumption of the system. 

To characterize the thermal efficiency of the Melter, three tests were run with varying changes. The PD feedback control system was used in each test, with a 50 $^{\circ}$C setpoint. We determined that the cartridge heaters consume a combined 69.96 W of power. We then calculated the lossless amount of heat required to raise the Melt Head from room temperature to 50 $^{\circ}$C. From this number and power consumption, we found the theoretical number of seconds that it would take to reach this temperature without losses. For each test, we measured the amount of time it took to reach 50 $^{\circ}$C, calculated the heat used, then compared this calculation to the lossless heat required to raise the Melt Head to 50 $^{\circ}$C. The first test was conducted without any thermal paste, the second with thermal paste between the Melt Head and Heating Plate, and the final test with thermal paste between the Melt Head and Heating Plate, as well as on each cartridge heater. As expected, the final test yielded the highest thermal efficiency with a value of 59.7\%.  

\begin{figure*} 
    \centering
    \includegraphics[width=7in]{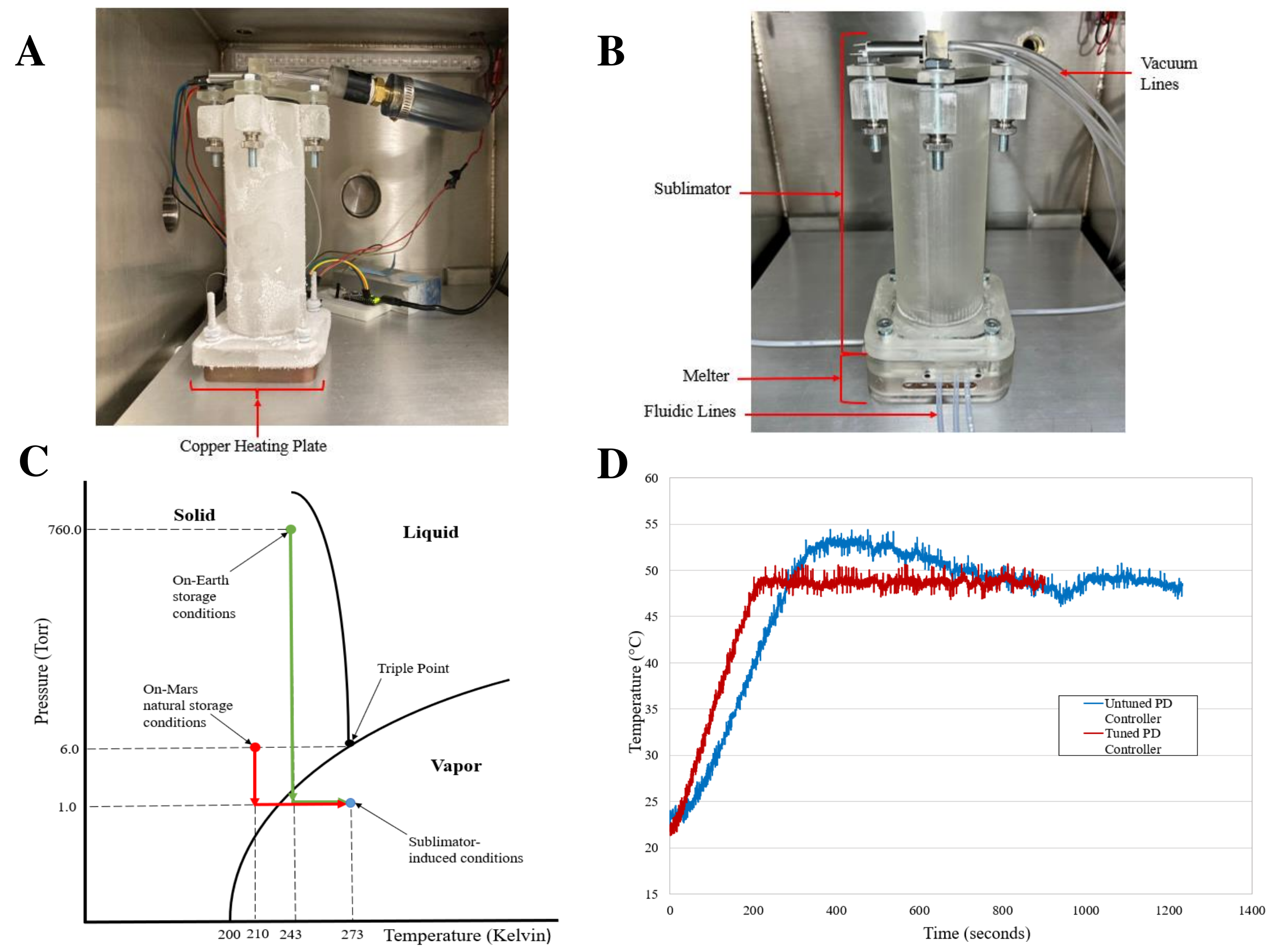}
    \caption{\bf{\label{fig_results}(A) Sublimator being tested inside thermal vacuum chamber, full melter not present. (B) Sublimator inside thermal vacuum chamber, melter present. (C) Pressure-Temperature diagram showing conditions induced by the Sublimator to begin the concentration sublimation process. (D) Tuned vs. untuned Proportional-Derivative (PD) temperature feedback results given a 50 $^{\circ}$C setpoint.}}
\end{figure*}

\section{Discussion}
\subsection{Limitations and Future Development}

The MSIS instrument has thus far demonstrated melting of a lab-made ice core sample in a benchtop setting. Upon the success of 3D printing the Melt Head, a future MSIS design could involve inner fluidic channels with more complex geometry, with outlet ports on a single side of the instrument; this modification would make for a more organized system in the lab and for future space missions. The next Melt Head design could include additional measures for preventing contamination of the inner core sections. For instance, the Melt Head partitioning walls could be made taller to ensure separation of the central core from the outer portion. Also, ultraviolet or infrared LEDs could be placed in a ring pattern at the outer base of the Melt Head to melt the contaminated outer region of the ice core before the inner regions, reducing the risk of meltwater mixing across Melt Head cavities. The next critical step in instrument development would be processing a Greenland or Antarctic ice core sample in the lab with MSIS, integrated with downstream analytical devices for processing validation. Upon such validation and necessary modifications, the MSIS instrument could begin utilization in the lab or in the field for Earth-based ice core studies.

At its current state, MSIS requires human operation and sample loading. A future vision is to make the MSIS system autonomous and compatible with a rover platform exploring ice deposits on other bodies. Automating a sublimator would require robust monitoring and data acquisition capabilities. MSIS currently relies on a user judging when enough sublimation has taken place based on the size of the sublimated ice sample. To process samples back-to-back, a cleaning process would be needed within its fluid lines. A one-size-fits-all approach to cleaning debris of different sizes would be challenging. However, the Sublimator component of MSIS does not limit itself to processing fully intact ice cores.

Considering the deposition of dust and debris on Mars, subsurface Martian ice has the potential to be stratified like permafrost ice deposits on Earth \cite{tr:mars-ice-2021}. The drilling process would inevitably accrue regolith in the ice cores which may pose a problem to the MSIS plumbing network. In addition, near-surface Mars ice could be more porous due to lower gravity and there may be a commensurate increase in the depth of pore close-off; this could lead to more fragile ice.

Rougher drilling techniques tend to produce ice “chips,” which are fragments defined to have a 6 mm maximum dimension \cite{tr:mars-ice-2021}. These chips lower the resolution and are difficult to decontaminate but are still considered useful for science.

Sublimation is an experimentally inefficient process \cite{Skelley2005} which does not pose an issue for benchtop experimentation on Earth but complicates its context in a space instrument. The convenient presence of low atmospheric pressure on Mars and other celestial bodies \cite{hartmuller2019demonstration} may allow for more efficient sublimation if it reduced or eliminated the need for a vacuum pump, as is needed on Earth. Without a vacuum pump, the sublimator instrument would save upwards of 1 kW of power and consume an estimated 76 W at its maximum power setting. Up to 300 W of power consumption in a space instrument is realistic for the capabilities of some rovers (e.g., Mars Oxygen ISRU Experiment, or MOXIE, instrument on Perseverance \cite{Hecht2021}) and serves as our metric when characterizing the MSIS power consumption.

\subsection{Integration with Future Human Lunar/Mars Missions}

The long-term objective is for future versions of MSIS to operate beyond Earth. With the Artemis III crew proposed to target regions near the Moon’s South Pole\footnote{\bf https://www.nasa.gov/press-release/nasa-identifies-candidate-regions-for-landing-next-americans-on-moon}, lunar water ice analyses are likely to take place. MSIS could contribute to Artemis III demonstrations of in-situ resource utilization for pre-processing lunar water ice for oxygen and hydrogen extraction to be used in fuel and life support systems. This would serve as a precursor to Mars missions with similar aims in processing ice to serve as an ingredient for propellant production while potentially searching for traces of life \cite{Spacek2021}. Also, in the context of preparing for the first human mission to Mars, potentially to mid-latitude regions in the 2030s, NASA has identified sampling subsurface ice as a critical science requirement for searching for life and investigating the evolution of Mars’ climate \cite{tr:mars-ice-2021}. If the Artemis III and Mars crews have the means to retrieve and manually handle ice core-like samples, MSIS—at its current state—could prove beneficial in preparing lunar water ice samples for in-situ refueling demonstrations, as well as helping meet NASA’s Mars human mission goals.

\subsection{Volcanism, Climate, and Solar Activity from Mars Ice}
An important caveat of our proposed analyses is that the formation and preservation mechanisms for different ice deposits on Mars is under study and without ground truth \cite{Williams2022}. At present, this means the availability of contiguous ice core records on Mars is unknown. This could complicate our ability to establish a high-resolution contiguous history from Mars ice.

Mars' obliquity, the angle between its orbital plane and its spin axis, is currently around 25$^{\circ}$, and it varies on a 120-thousand-year (Ky) timescale, with the amplitude of this cycling varying chaotically on ~1.2 My timescales \cite{Touma1993,Ward1973}. During periods of high obliquity, polar ice is mobilized and falls as snow in mid-latitude and equatorial regions \cite{WEISS2019115847}. This process has resulted in repeated mid-latitude “ice ages”  for an estimated total of ~680 My out of the last 3.6 billion years (Gy). Thus, it is reasonable to assume that current remnants of these ice ages are mainly locally-contiguous deposits, even if a global contiguous record is not expected.

\subsubsection{Detection of Recent Volcanism on Mars}

Mars has been geologically active until the recent past, and while no active volcanism has been observed, it is unknown when it ceased. Ice cores on Earth are typically synchronized using volcanic events. Emissions of \ch{SO2} are oxidized to sulfuric acid and measured in ice cores as acidity and sulfate (\ch{SO4^2-}) pulses above the climate background \cite{RASMUSSEN200818}. The present-day atmosphere of Mars is highly oxidizing and has been so for billions of years, except during brief warmer and wetter reducing periods \cite{Wordsworth2021}. Thus, volcanism on Mars could leave signatures of volcanic events within Mars ice similar to those we analyze on Earth (e.g., \cite{Settle1979,CRADDOCK2009512}).

\subsubsection{Characterization of Mars Climate in the Late Amazonian}
Mars has been largely dry, cold, and oxidizing for the last 3 Gy, a period denoted as the Amazonian. Ice core records on Mars, potentially going back tens to hundreds of My \cite{Bramson2017,Levy2021}, can help establish recent variability of the climate during this period. While the exquisite layering of the Mars polar ice caps contains important climate information, these deposits may only date to the last ~4 My \cite{Levrard2007}. While young, polar deposits may offer a more contiguous, although dusty, record during this recent period compared to mid-latitude ice deposits. However, these deposits are much more difficult to access due to orbital mechanics and harsher conditions, making mid-latitude ice a compelling target.

\subsubsection{Seeking Evidence of Subsurface Life}
Life on Mars, if it exists, could be related to Earth. A Mars origin for Earth life could resolve significant inconsistencies between the inferred history of life and Earth's geologic history \cite{Carr2022}. If life got started on Mars, there is little reason to expect it is now extinct, given the significant overlap between the temperature and pressure regimes occupied by life on Earth and those available on Mars \cite{Jones2011}. In short, the most likely place for life on Mars today is in the deep subsurface. Given this, and that current (e.g., Curiosity, Perseverance) and future (e.g., Rosalind Franklin) missions are limited to cm to m drill depths, how could we sample the deep subsurface in the near term, without the mass required to drill to km-scale depths?

On Earth, phreatomagmatic eruptions, those involving magma interacting with water, yield explosive events, which can result in the dispersal of even large microorganisms such as diatoms up to nearly 1000 km \cite{VanEaton2013}. Evidence of such eruptions has been identified on Mars, representing interactions between magma and ground ice. For example, explosive volcanism in Elysium Planitia \cite{CASSANELLI201896} may have occurred as recently as 46 to 222 Ky ago \cite{HORVATH2021114499,MOITRA2021116986}. This region, home to the InSight Lander, could thus possibly harbor active volcanism, which could offer an abode for life. 

Therefore, explosive eruptions, as well as large impact events, could eject life or signs of life from deeper within the Martian crust, and their remnants could be deposited on accumulating ice sheets during periods of high obliquity. Because such explosive events are occurring up to the present day, geologically speaking, such events have likely been ongoing over the entire history of Mars mid-latitude ice emplacement. Burial within accumulating ice deposits would enhance the preservation of organics relative to long-term exposure at the surface, due to shielding from space radiation and a lack of thermal cycling. Such remnants of life, if present, would likely be present at extremely low abundance. However, they could be concentrated, for example as described by \cite{Spacek2021} during large-scale processing of ice for propellant production.

\subsubsection{Returned Samples for Cosmogenic Nuclide Analyses}
Cosmogenic nuclides are isotopes produced by interactions between atoms and high energy galactic cosmic rays (GCR). On Earth, Beryllium-10 (\isotope[10]{Be}) is used to infer historical solar activity from its abundance in ice cores. Cosmogenic radionuclides can also reveal extreme solar events \cite{Paleari2022} and nearby supernovas \cite{PhysRevLett.125.031101}.

On Earth, once \isotope[10]{Be} forms and attaches to aerosols, it takes 1 to 2 years for deposition to occur \cite{Zheng2020}. Similar modeling of transport and scavenging processes, as well as important baseline measurements, would be required for Mars to interpret a record of cosmogenic isotopes recovered from Mars ice \cite{DORAN2004313}. Recently, the potential for \isotope[10]{Be} analysis from ice chips, permitting much faster drilling, has been demonstrated \cite{NGUYEN2021100012}.

While the 1.39 My half-life of \isotope[10]{Be} permits its use through the full range of ice ages on Earth, alternative cosmogenic isotopes would be needed for older ice on Mars. Candidates include \isotope[53]{Mn}, with a half-life of 3.7 My, and \isotope[129]{I}, with a half-life of 15.7 My. 

\isotope[53]{Mn} can be produced within supernova events, and has been detected co-located with \isotope[60]{Fe}, also produced during supernovas, within marine sediments. This has been used to date supernova events \cite{PhysRevLett.125.031101}. \isotope[53]{Mn} is also produced terrestrially, through cosmic ray bombardment and in-situ production \cite{Schaefer2006}.

The only important target element for \isotope[53]{Mn} production in situ is iron \cite{Schaefer2006}, which is abundant in the dust that undergoes local to global aeolian transport on Mars. Dust aloft might experience higher GCR rates due to a lack of atmospheric shielding, although this effect could be marginal due to the relatively thin atmosphere of Mars (1\% of Earth in the geologically recent past). Dust deposited on ice sheets would continue experiencing cosmogenic production of \isotope[53]{Mn} during initial burial. At the Greenland summit (e.g., GISP2 core), historical accumulation rates are from 15 to 40 cm/y ice equivalent. On Mars, accumulation rates are likely much lower, perhaps 10 mm/y \cite{MADELEINE2009390}, but are not well constrained. This means isotopic signals on Mars might be spread out over broader ($>$ annual) periods than on Earth.

For even older ice, \isotope[129]{I} analysis may be useful. \isotope[129]{I} is mainly produced by the fission of uranium, with production by cosmic rays also contributing. On Earth, production is estimated to be $10^{-14}$ grams of \isotope[129]{I} per gram of \isotope[127]{I} \cite{Edwards1962}. It has been proposed that Mars may have had a naturally-occurring fission reactor in the Mare Acidalium region \cite{Brandenburg2011}. If true, this could have resulted in an explosion on the order of 1 Gy ago, which would have scattered radioactive iodine across the planet’s surface. This could make it harder to attribute radioactive iodine to cosmic ray sources.

The most sensitive approach known to measure \isotope[53]{Mn} and \isotope[129]{I} is accelerator mass spectrometry (AMS) \cite{DONG201558,Takashi2006}. Unfortunately, AMS facilities are large and complex and would be unlikely candidates for in-situ analysis. Instead, ice could be processed in situ, sublimated, with retention of isotopes bound to particulates, and those residues could be returned for AMS analysis on Earth. Searching for rare isotopes might require processing large volumes of ice; for example, 500 kg of Antarctic ice was analyzed to identify interstellar \isotope[60]{Fe} by AMS \cite{PhysRevLett.123.072701}.

\subsection{Enhancing Limit of Detection for Ocean Worlds}
Based on analysis of nutrient-limited Earth environments, modeling of energy limitations, and other factors, estimates for cell abundances at Ocean Worlds such as Enceladus and Europa may be as low or lower than 1000 to 100 cells/ml
\cite{MacKenzie2022}. Most instrumentation flown or under development may not have adequate analytical limits of detection without concentration. Concentration approaches include filtration, flow focusing, and sublimation, among others. A prototype sublimator was previously developed and was able to demonstrate the freeze-drying of a 5 ml sample in 5 hours at under 5 W power \cite{hartmuller2019demonstration}. This was achieved by accounting for the available cold temperatures and low pressure expected to be available under Europa or Enceladus surface conditions. Even so, sublimating large volumes of ice is likely infeasible due to the high energy cost of sublimation; we can expect to have higher energy budgets on Mars where higher power operation could process larger volumes via sublimation. Cold and low-pressure conditions are also available on Mars, where surface pressure is ~600 to 700 Pa; a vacuum pump will be required to maintain pressure below the triple point of water, ~610 Pa.

\section{Conclusions}
From processing true ice cores to automating some capabilities, there is much more subsystem development to occur before MSIS is used with human or robotic missions to Mars and icy bodies. Fortunately, MSIS has equally diverse applications on Earth and its functionalities can be verified with Earth-based ice samples. In the long-term, MSIS would greatly simplify the processes needed for ice science in other planetary environments. Our initial work toward a space-mission-compatible ice core processing instrument could someday enable the examination of extraterrestrial ice cores to help address fundamental and diverse questions in planetary processes, habitability, astrobiology, and heliophysics.



\acknowledgments
Funded by faculty startup funds from the Georgia Institute of Technology to C.E.C.

\bibliographystyle{IEEEtran}
\newcommand\BIBentryALTinterwordstretchfactor{1}
\bibliography{Bibliography}

\thebiography
\begin{biographywithpic}
{Alexander G. Chipps}{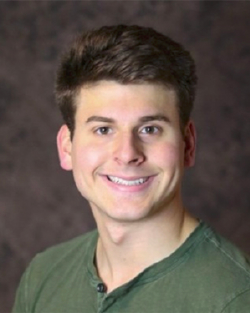} is an undergraduate researcher at Georgia Tech, set to graduate with a B.S. degree in Mechanical Engineering in 2023. He aims to partake in graduate Aerospace Engineering research to further develop space instrumentation for life detection.
\\
\end{biographywithpic} 

\begin{biographywithpic}
{Cassius B. Tunis}{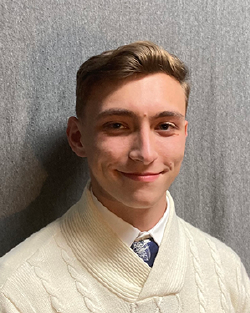}
is an undergraduate researcher at Georgia Tech pursuing a B.S. degree in Aerospace Engineering with a minor in Earth and Atmospheric Science. He is interested in building instruments to better understand planetary processes.
\\
\\
\end{biographywithpic}

\begin{biographywithpic}
{Nathan Chellman}{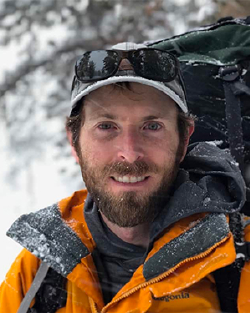} holds a B.Sc. in Geology/Biology from Brown University, and M.S. and Ph.D. degrees in Hydrology from the University of Nevada, Reno. He is an Assistant Research Professor and snow and ice hydrologist in the Division of Hydrologic Sciences at the Desert Research Institute (DRI). He specializes in the collection, processing and analysis of ice cores.
\end{biographywithpic}

\begin{biographywithpic}
{Joseph R. McConnell}{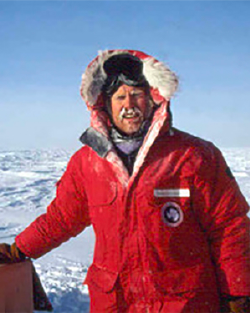} received a B.A. in Geology and Geophysics from Yale, an M.S. in Exploration Geophysics from Stanford University, and a Ph.D. in Hydrogeology and Water Resources from the University of Arizona. He is a Research Professor in the Division of Hydrologic Sciences at the Desert Research Institute (DRI). Current NSF, NASA, and internally funded research projects include ice core chemistry-based studies in Greenland, Antarctica, and the Americas, using his group's unique ultra-trace ice core analytical laboratory.
\end{biographywithpic}

\begin{biographywithpic}
{Bruce Hammer}{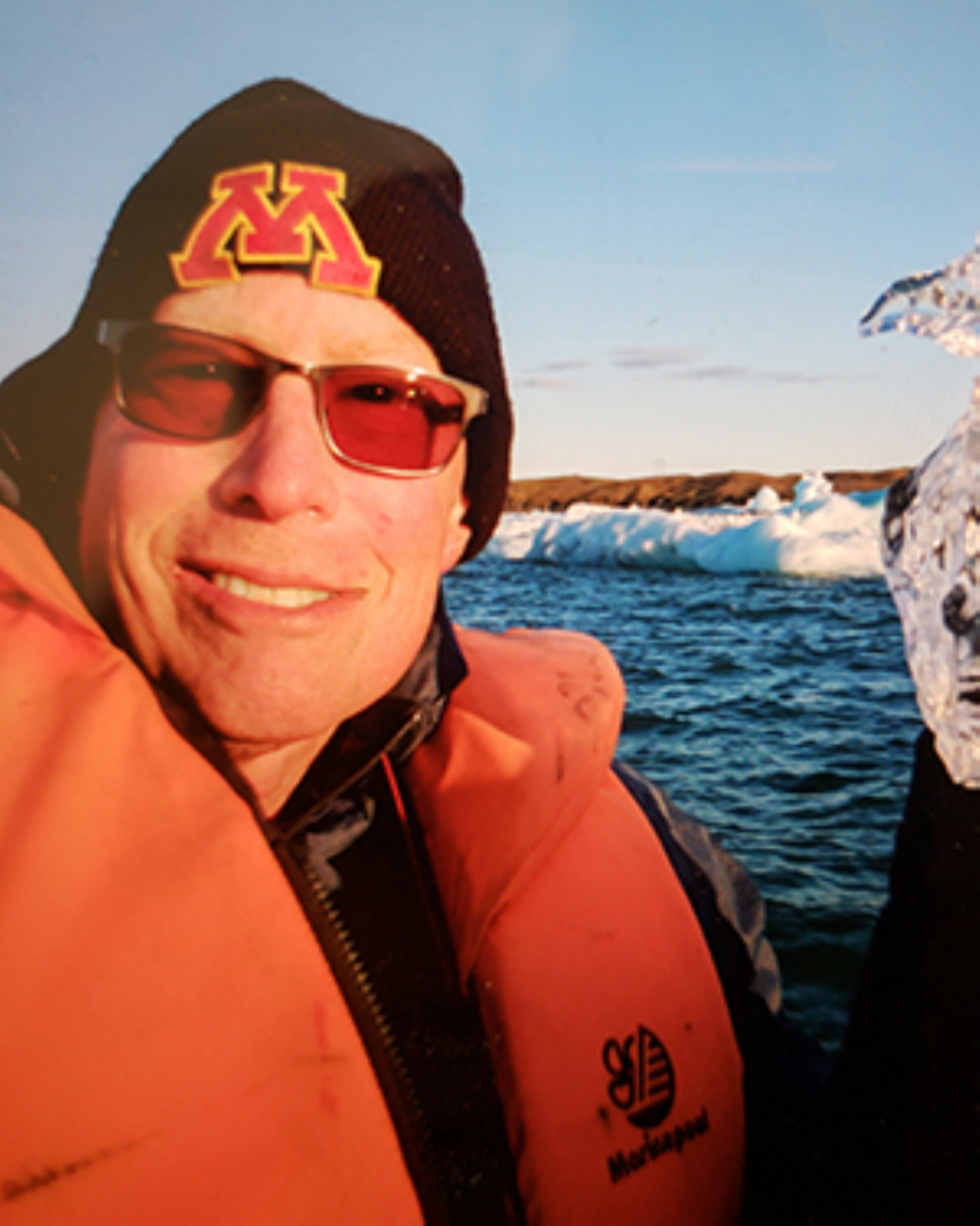} received the B.S. in Biology\/Physics from the State University of New York in 1974, the M.S. in Biomedical Engineering from Northwestern University in 1976, and the Ph.D. in Biomedical Engineering from Northwestern University in 1981. He is a Professor of Radiology \& Adjunct Professor of Entrepreneurial Studies at the University of Minnesota. His work focuses on hardware development and its implementation in support of biomedical and astrobiology applications, including a NASA-funded sublimator device for Ocean Worlds.
\end{biographywithpic}

\begin{biographywithpic}
{Christopher E. Carr}{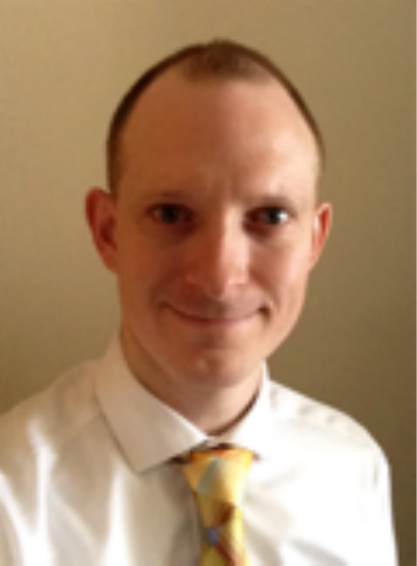} received his B.S. degree in Aero/Astro and Electrical Engineering in 1999, his Masters degree in Aero/Astro in 2001, and his Sc.D. degree in Medical Physics in 2005, all from MIT. He is an Assistant Professor in the Daniel Guggenheim School of Aerospace Engineering and the School of Earth and Atmospheric Sciences at the Georgia Institute of Technology. He is broadly interested in searching for and expanding the presence of life beyond Earth.
\end{biographywithpic}

\end{document}